\documentclass[letter]{aa}  
\usepackage{natbib}
\bibpunct{(}{)}{;}{a}{}{,} 

\usepackage{graphicx}
\usepackage{txfonts}
\usepackage{natbib,twoopt}
\usepackage[breaklinks=true]{hyperref} 
\bibpunct{(}{)}{;}{a}{}{,}             
\makeatletter
  \newcommandtwoopt{\citeads}[3][][]{\href{http://adsabs.harvard.edu/abs/#3}%
    {\def\hyper@linkstart##1##2{}%
     \let\hyper@linkend\@empty\citealp[#1][#2]{#3}}}
  \newcommandtwoopt{\citepads}[3][][]{\href{http://adsabs.harvard.edu/abs/#3}%
    {\def\hyper@linkstart##1##2{}%
     \let\hyper@linkend\@empty\citep[#1][#2]{#3}}}
  \newcommandtwoopt{\citetads}[3][][]{\href{http://adsabs.harvard.edu/abs/#3}%
    {\def\hyper@linkstart##1##2{}%
     \let\hyper@linkend\@empty\citet[#1][#2]{#3}}}
  \newcommandtwoopt{\citeyearads}[3][][]%
    {\href{http://adsabs.harvard.edu/abs/#3}
    {\def\hyper@linkstart##1##2{}%
     \let\hyper@linkend\@empty\citeyear[#1][#2]{#3}}}
\makeatother

\usepackage{subfigure}

\def\sigsfr{$\Sigma_{\rm SFR}$ }
\def\sigsfrend{$\Sigma_{\rm SFR}$}
\def\Mtot{$\textrm{M}_{\star}$ }
\def\Mtotend{$\textrm{M}_{\star}$}
\def\sigmass{$\Sigma_{\star}$ }
\def\sigmassend{$\Sigma_{\star}$}
\def\dsigsfr{$\Delta \Sigma_{\rm SFR}$ }
\def\dsigsfrend{$\Delta \Sigma_{\rm SFR}$}

\def\pipedend{{\tt Pipe3d}}
\def\piped{{\tt Pipe3d} }

\begin{document}

   \title{Are all starbursts equal?  Star-formation-rate profiles in merger versus secular starbursts}
   \titlerunning{Star formation rate profiles in merger vs. secular starbursts}

   \author{Mallory D. Thorp
          \inst{1},
          Sara L. Ellison\inst{2},
          \and
          Ana Galicia\inst{2}
          }

   \institute{Argelander-Institut für Astronomie, Universität Bonn, Auf dem Hügel 71, 53121 Bonn, Germany\\
              \email{mthorp@uni-bonn.de}
         \and
             Department of Physics \& Astronomy, University of Victoria, Finnerty Road, Victoria, British Columbia, V8P 1A1, Canada\\
             }

   \date{Received July 26, 2024; accepted September 13, 2024}

 
  \abstract
   {} 
   {Galaxy interactions can trigger drastic changes in the resolved star-forming properties of their constituents, but it remains unclear as to whether those changes are discernible from secular starburst triggers. In this Letter we investigate whether or not post-merger galaxies create unique star-forming trends on a kiloparsec scale.}
   {We present radial trends in star-formation-rate (SFR) surface density ($\Sigma_{\mathrm{SFR}}$) for 150 post-merger galaxies with moderate to extremely heightened global SFRs using observations from the Mapping Nearby Galaxies at Apache Point Observatory (MaNGA) survey. We juxtapose these profiles with those of noninteracting galaxies (excluding both galaxy pairs and post-merger galaxies) with similarly enhanced global SFRs.}
   {Post-merger galaxies have a much stronger central starburst than isolated galaxies with similar global star-formation enhancements. Indeed, isolated starburst galaxies (SBs) lack a marked central enhancement and instead show a fairly uniform enhancement in $\Sigma_{\mathrm{SFR}}$ with radius. Moreover, the difference in central star formation between post-merger galaxies and noninteracting galaxies is more radially extended and pronounced when the global enhancement in star formation is larger. We conclude that post-merger galaxies create a unique signature in their resolved star-forming properties that is distinct from secular processes that can trigger similar global SFR enhancements.}
   {}

   \maketitle
%

\section{Introduction}

Both theoretical analysis and observational evidence indicate that galaxy mergers can have a profound impact on the star formation activity within a galaxy. The gravitational torques that occur as two galaxies interact cause gas to lose angular momentum and fall toward the galactic center \citep{Barnes1991FuelingMergers,Mihos1994DenseRemnants,Mihos1996GasdynamicsMergers,Iono2004RadialObservations,Hopkins2009TheDemographics}. These gaseous inflows fuel a subsequent burst of star formation activity in the center of the galaxy that is strongest near coalescence \citep{Springel2000ModellingGalaxies,DiMatteo2005EnergyGalaxies,Montuori2010TheMergers,Torrey2012THEGALAXIES,Hopkins2013StarMedium,Moreno2019InteractingContent, Patton2020InteractingContext}. Observations have confirmed this theoretical framework; on average, global star formation rates (SFRs) are enhanced as the distance between galaxy pairs diminishes \citep{Barton2000TidallyGalaxies,Nikolic2004StarSurvey,Ellison2008GalaxyRelation,Woods2010TRIGGERED0.08-0.38,Scudder2012GalaxyKpc,Patton2016GalaxySeparations}, and single galaxies with highly asymmetric morphologies indicative of post-merger galaxies exhibit the strongest enhancement in SFR \citep{Ellison2013GalaxyGalaxies,Bickley2022StarUNIONS}. This burst of star formation may be followed by rapid quenching \citep{Pawlik2016ShapeStages,Rowlands2018SDSS-IVProperties,Wilkinson2022TheUNIONS,Ellison2022GalaxyFormation,Li2023Post-starburstGalaxies}.

The recent prominence of large integral field spectroscopic (IFS) surveys has enabled the expansion of  the exploration of merger-induced changes in star formation to kiloparsec-scale variations. IFS observations of hundreds of galaxy pairs have shown that interaction-driven star formation is not limited to the center of galaxies, particularly in the late stages of an interaction \citep{Pan2019SDSS-IVInteractions,Thorp2019SpatiallyMaNGA, Thorp2022TheFormation}. However, the strength of the central starburst  does correlate with the progression of the interaction, as pairs with small separations and clear morphological signs of an interaction tend to have heightened central star formation activity compared to both other interacting galaxies and a noninteracting control sample \citep{Pan2019SDSS-IVInteractions,Steffen2021}. The radial distribution and strength of star formation activity is also dependent on properties beyond the interaction stage: specifically the mass ratio between constituents \citep{Steffen2021}. These works have demonstrated how late-stage mergers have SFR distributions that are  distinct from those of isolated galaxies on average. However, they have not investigated whether these changes can be distinguished from secular starburst galaxies (SBs, for brevity), which themselves have star-formation gradients distinct from those of galaxies with moderate or suppressed global SFRs \citep{Ellison2018StarMaNGA,Sanchez2020SpatiallyGalaxies}.

Despite this clear connection between mergers and SBs, elevated SFRs can be caused by other processes as well. For example, disk instabilities may arise independently of minor or major mergers, resulting in a comparable infall of gas and a centrally concentrated burst of star formation \citep{Tacchella2016TheReplenishment}. Recent results from COSMOS-Web have revealed that, though there is an increased fraction of mergers in SB samples out to a redshift of $4$, isolated disk galaxies constitute a significant fraction of SBs, often with clumpy star formation characteristic of disk instabilities \citep{Faisst2024COSMOS-Web:Z4}. Simulations demonstrate how stellar bars can also transport gas to fuel a central starburst \citep{Athanassoula1992TheBars,Athanassoula1994GasPaper}, and observations confirm that barred galaxies often have higher nuclear star formation activity than their unbarred counterparts \citep{Hawarden1986EnhancedGalaxies,Ho1997TheActivity,Ellison2011GalaxyNuclei,Oh2012BARACTIVITY,Lin2017SDSS-IVMaNGA,Chown2019LinkingGalaxies}. Understanding these processes in conjunction with mergers is crucial for untangling the various forces that influence resolved star-formation evolution in observational extragalactic studies.

In this Letter, we present the first statistical study focused on how the resolved star-formation properties of mergers differ from those of isolated galaxies with similar global SFR enhancements. We focus on the post-merger stage in order to negate any differences that may result from the progression of an interaction. Using the Mapping Nearby Galaxies at Apache Point Observatory (MaNGA) IFS survey, we are able to collect the largest sample of post-merger galaxies with IFS observations to date, along with specific and well-matched control galaxies not currently undergoing an interaction. In Section \ref{Sec: Methods} we outline how the merger sample and viable comparison galaxies are selected, as well as how we quantify enhancements in star formation activity. We then present the resolved difference in star formation between post-merger galaxies and similar isolated galaxies in Section \ref{Sec: Results}, followed by a brief discussion and final remarks in Section \ref{Sec: Discussion}. We adopt a cosmology in which $\textrm{H}_0=70$km/s/Mpc, $\Omega_{\textrm{M}}=0.3$, and $\Omega_{\Lambda}=0.7$, and a Salpeter initial mass function \citep{Salpeter1955THEEVOLUTION}.

\section{Methods}
\label{Sec: Methods}
Given both the rarity of post-merger galaxies in a galaxy population and the anticipated range of transformation resulting from an interaction, a large sample of galaxies is required to capture the breadth of star-formation behaviors in the aftermath of mergers. We elect to use the MaNGA survey, the largest IFS survey to date with 10,010 unique targets observed \citep{Bundy2015OVERVIEWOBSERVATORY,Law2021SDSS-IVAccuracy,Abdurrouf2022TheData}. MaNGA employs hexagonal bundles of 2'' spectroscopic fibers, ranging in sizes from 12.5'' to 32.5'' such that each galaxy is observed out to at least 1.5 effective radii \citep{Law2015ObservingSurvey}. As such, this survey allows us to collect a large number of both post-merger and isolated galaxies with kiloparsec(kpc)-scale SFR maps covering a majority of the disk. The full MaNGA sample is selected from the NASA-Sloan Atlas (NSA, \citealt{Blanton2011ImprovedImages}), which also provides key galaxy properties used in this analysis, including redshift ($z$), effective radius ($R_e$), and $b/a$ axial ratio.

The spatially resolved data products used in this work are derived from \pipedend, a spectroscopic analysis pipeline that uses the single stellar population (SSP) analysis tool {\tt fit3d}\footnote{http://www.astroscu.unam.mx/sfsanchez/FIT3D} to provide information on both stellar population and emission line fits \citep{Sanchez2016CALIFARelease,Sanchez2016Pipe3DFIT3D,Sanchez2022SDSS-IVGalaxies}. For this analysis, we adopt the \piped dust-corrected stellar-mass surface densities (\sigmassend), as well as H$\alpha$, H$\beta$, [OIII]$\lambda$5007, and [NII]$\lambda$6584 fluxes, all of which are necessary to compute accurate SFR surface densities (\sigsfrend). \sigsfr are derived from the H$\alpha$ flux, which is corrected for reddening effects, assuming a fixed H$\alpha$/H$\beta$ of 2.86 and a Milky Way extinction curve \citep{Cardelli1989THEEXTINCTION}. The dust-corrected H$\alpha$ luminosity is multiplied by $7.9\times10^{-42}$ to determine \sigsfr in units of $\mathrm{M}_{\odot} \mathrm{yr}^{-1} \mathrm{kpc}^{-2}$ \citep{Kennicutt1994PastGalaxies}. Both \sigmass and \sigsfr are inclination corrected using the $b/a$ axial ratio provided in the NSA catalog.

To ensure that the SFR surface density derived from H$\alpha$ luminosity stems from recent star formation, and not active galactic nuclei or low-ionization emission regions, we limit our spaxel sample to those that meet the \cite{Kauffmann2003TheNuclei} star-forming criterion. We also require a signal-to-noise ratio of greater than 3 for all emission lines required in that diagnostic (H$\alpha$, H$\beta$, [OIII]$\lambda$5007, and [NII]$\lambda$6584). To further minimize contamination from older stellar populations, we additionally require the H$\alpha$ equivalent width be greater than 3\AA \citep{CidFernandes2011AAGN}.

Each spaxel is assigned an inclination-corrected galactocentric radius value in units of $R_e$, the half-light radius determined from the single-component Sérsic fits from the NSA catalog. $R_e=0$ is located at the center of the integral field unit (IFU), unless the maximum V-band flux is more than 10\% of the total IFU width away from the IFU center (in this latter case, the maximum V-band flux is used as the true center of the galaxy). Thus, when we construct radial profiles of spaxel properties, we can compare similar locations in each galaxy despite having galaxies of a variety of sizes.

Integrated properties derived from \piped are also provided as part of the most recent MaNGA data release \citep{Lacerda2022PyFIT3DPipeline,Sanchez2022SDSS-IVGalaxies}. From this value-added catalog, we collect the total stellar mass (\Mtotend) and total SFR of each galaxy, as both are crucial in our comparison of merger SBs with secular SBs as described in Section \ref{Sec: Merger Sample}.

A final global property we compute for this analysis is the offset from the global star-forming main sequence ($\Delta\mathrm{SFR}$), which we use to identify SBs triggered by both mergers and secular processes. Star-forming galaxies exhibit a strong correlation between \Mtot and SFR, which is often called the star-forming main sequence (SFMS, \citealt{Brinchmann2004TheUniverse,Noeske2007StarGalaxies,Daddi2007MultiwavelengthGrowth,Schreiber2015TheDay}). $\Delta\mathrm{SFR}$ measures the difference between a single galaxy's global SFR and the median SFR of galaxies of the same stellar mass on the SFMS (MS galaxies hereafter). For this analysis, MS galaxies are those with global SFR and \Mtot that are within 2$\sigma$ of a first-order polynomial fit to the SFMS of all MaNGA galaxies. $\Delta\mathrm{SFR}$ is then computed for each galaxy by taking the difference between the target galaxy SFR and the median SFR of all MS galaxies matched within 0.1 dex in \Mtot and 0.005 in $z$ to the target galaxy. Using a median SFR value rather than the SFR established by the SFMS fit allows us to control for variations in SFR with redshift and limit the impact of the specific SFMS fit (which can vary with different samples and fitting techniques).

\subsection{Spaxel SFR offsets}
\label{Sec: Sample}

Rather than compare \sigsfr distributions for our two samples directly, we elect instead to analyze profiles of the offsets from the resolved star-forming main sequence (rSFMS, \citealt{Cano-Diaz2016SPATIALLYSURVEY, GonzalezDelgado2016StarSequence, Hsieh2017SDSS-IVSequence,Medling2018TheMain}). Just as there is a strong correlation between SFR and \Mtot for star-forming galaxies, there exists a correlation between \sigsfr and \sigmass 
on a kpc scale that can define ``normal'' star-forming behavior for an individual spaxel. We define an offset from the rSFMS (\dsigsfrend) similarly to previous MaNGA studies (see \citealt{Ellison2018StarMaNGA,Pan2019SDSS-IVInteractions,Thorp2019SpatiallyMaNGA}):
\begin{equation}
\Delta \Sigma_{\mathrm{SFR}}=\Sigma_{\mathrm{SFR}}^{\textrm{target}}-<\Sigma_{\mathrm{SFR}}^{\textrm{control}}>,
\end{equation}
\noindent where $<\Sigma_{\mathrm{SFR}}^{\textrm{control}}>$ is the median \sigsfr value for a set of control spaxels within 0.1 dex of \sigmassend, 0.1 dex of \Mtot, and 0.1$R_e$ of the radius value of the target spaxel. The control spaxels are also limited to those in galaxies with inclinations of less than 70\textdegree, and that are not a currently merging or post-merger system (see Section \ref{Sec: Merger Sample} for merger classification). We note that though control spaxels for this calculation are selected from only isolated galaxies in MaNGA, this is a separate process from the selection of individual isolated SB \dsigsfr profiles compared to post-merger galaxies.

\subsection{Post-merger and isolated starburst selection}
\label{Sec: Merger Sample}

One co-author  visually identified  400 post-merger galaxies  out of the 10,010 MaNGA targets by examining {\it gri}-images from the Sloan Digital Sky Survey (SDSS). Galaxies that have recently undergone an interaction will have a range of highly to moderately disturbed morphologies, in particular tidal tails and asymmetric shells. If there is a second galaxy either on the IFU or with a visible connection to the galaxy, we exclude these from the analysis as they are likely late-stage interacting pairs. To minimize the number of targets that resemble a post-merger system but have a companion beyond the immediate field of view, we also exclude galaxies from the spectroscopic pair catalog provided by \cite{Patton2016GalaxySeparations}. Any post-merger galaxy that has a companion in this catalog closer than 100 kpc in projected separation and 500 km/s in velocity space is excluded from the analysis. We further exclude galaxy pairs identified with this catalog or from visual inspection from the pool of MaNGA galaxies considered for the isolated SB sample. Lastly, we limit our analysis to post-merger galaxies experiencing some amount of star-formation enhancement, that is, $\Delta\mathrm{SFR}>0$ dex. Out of 400 initial post-merger galaxies, only 206 have $\Delta\mathrm{SFR}>0$ dex. A summary of the post-merger selection process is provided in Appendix
\ref{app_merger}, along with details of the interacting galaxy pairs removed from this analysis.

To discern whether mergers create unique changes in kpc-scale star formation, each merger must be compared to similar but isolated SB. Though there are a variety of properties with which to characterize galaxies, we choose ``similar'' galaxies based on \Mtot and $z$ to best constrain variations on the rSFMS. \Mtot is one of the key factors in regulating a galaxy's star formation activity \citep{Brinchmann2004TheUniverse, Salim2007UVUniverse, Pannella2009STARDOWNSIZING, Renzini2015AnGalaxies}, and given that global and local star-formation relations are closely linked, it is crucial to only compare post-merger galaxies to noninteracting galaxies with a similar \Mtot \citep{Rosales-Ortega2012ARELATION,Sanchez2013Mass-metallicityRate,Cano-Diaz2016SPATIALLYSURVEY,Hsieh2017SDSS-IVSequence}. The normalization of the SFMS also evolves with $z$, with higher SFRs at fixed stellar mass \citep{Daddi2007MultiwavelengthGrowth,Noeske2007StarGalaxies,Schreiber2015TheDay}. Thus, in order to make a fair comparison, we must constrain the redshift of the control galaxies as well. Finally, we match our post-merger galaxies to controls with similar offsets from the SFMS ($\Delta\mathrm{SFR}$, as previously described).

We assign post-merger galaxies a subsample of comparison isolated SBs\footnote{For simplicity we use the term ``starburst'' when referring to the isolated galaxy sample matched for fair comparison to the post-merger sample, but these galaxies can have any value of $\Delta\mathrm{SFR}>0$ dex. Thus many ``starburst'' with low $\Delta\mathrm{SFR}$ will fall within the SFMS.} matched on $z$($\pm0.005$), \Mtotend($\pm$0.1dex), and $\Delta\mathrm{SFR}$($\pm$0.1dex), and require that each post-merger system have at least five isolated SBs for comparison. Of the 206 post-merger systems considered, 56 have less than five viable isolated SB controls. Further discussion of why these post-merger galaxies fail to acquire the minimum number of isolated SB controls is provided in Appendix \ref{app_control}.

\begin{figure}
    \includegraphics[width=0.99\columnwidth]{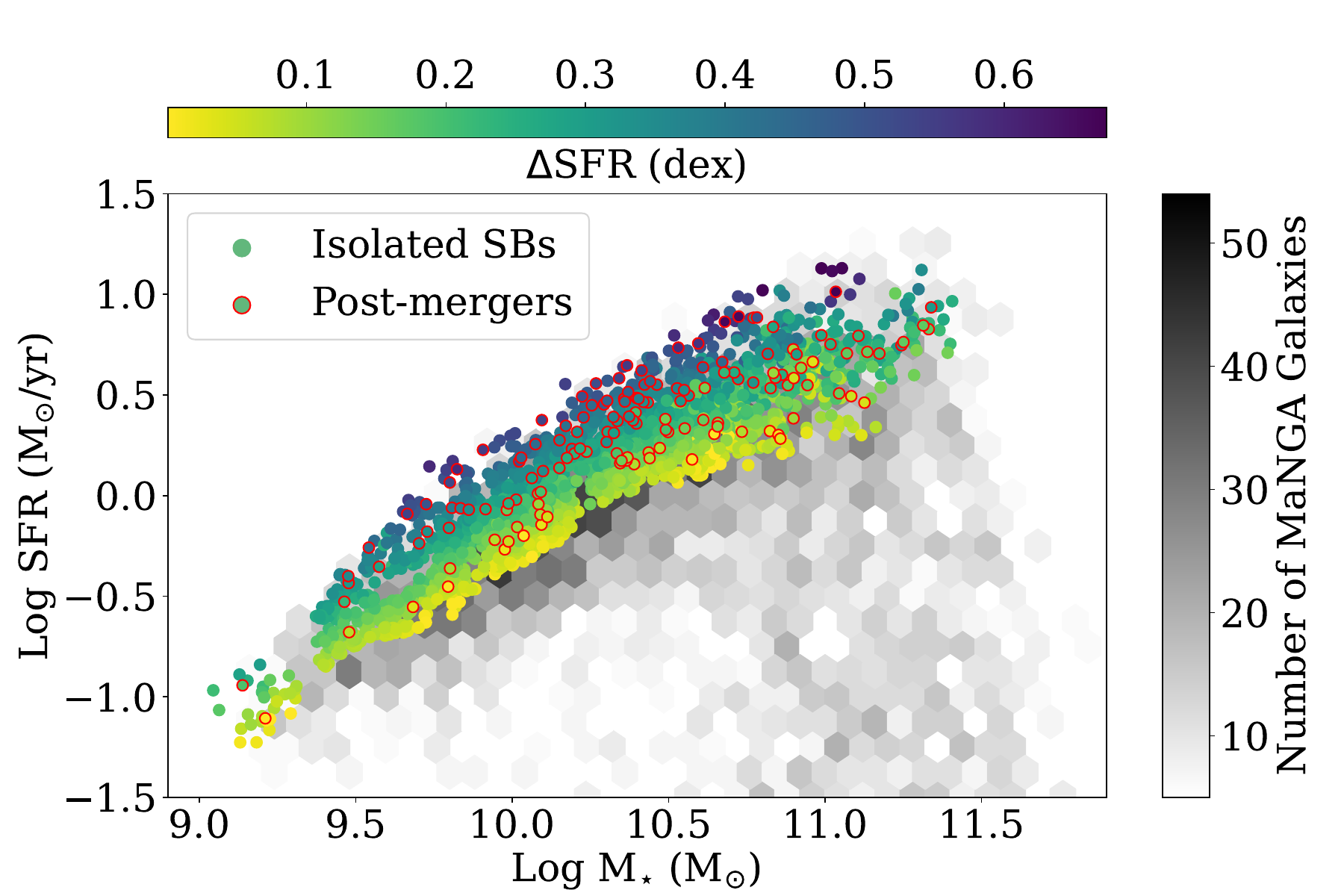}
    \centering
    \caption{ SFR vs. \Mtot of each post-merger galaxies (colored points circled in red) and the isolated SBs matched for comparison (borderless colored points). Points are color-coded according to the offset from the SFMS ($\Delta\mathrm{SFR}$); many post-merger galaxies and isolated SBs have small enough $\Delta\mathrm{SFR}$ that they overlap with the SFMS (though $\Delta\mathrm{SFR}>0$ dex for all colored points). A density histogram of the full MaNGA sample is shown in gray in the background.}
    \label{fig:SFMS}
\end{figure}

The final sample consists of 150 post-merger galaxies with $\Delta\mathrm{SFR}>0$ dex, and 1676 isolated SBs used as comparison galaxies. These two samples are shown with respect to the SFMS in MaNGA in Figure \ref{fig:SFMS}, with the post-merger galaxies circled in red. Though the most extreme SBs are removed due to a lack of well-matched control galaxies, both the post-merger and isolated SB sample still reach significant elevation above the SFMS.

\begin{figure*}
    \includegraphics[width=0.99\textwidth]{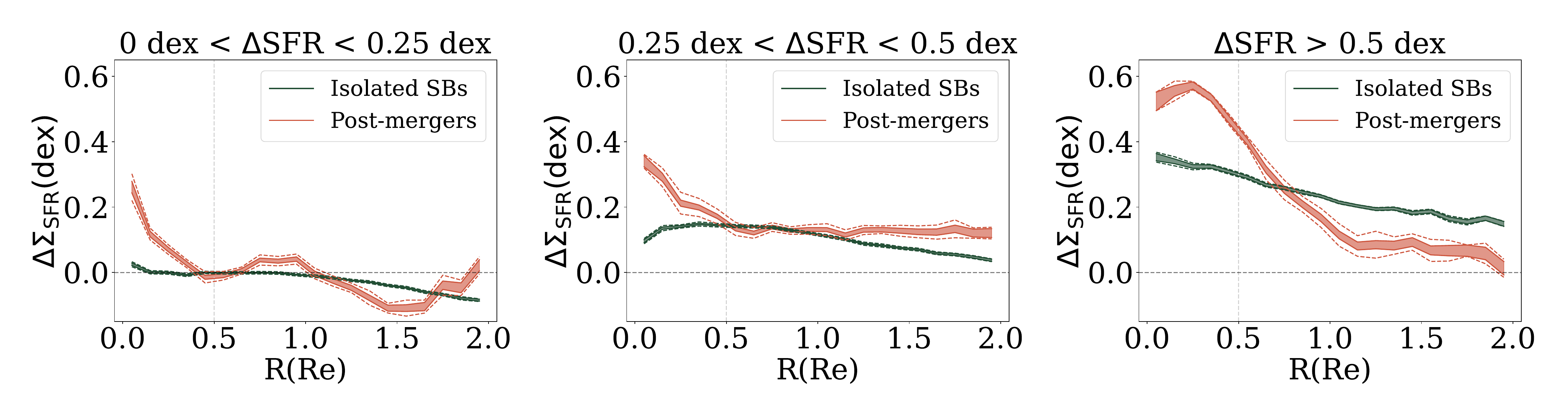}
    \centering
    \caption{Profiles of the median \dsigsfr value in 0.1$R_e$ radial bins for the post-merger galaxies (red) and matched isolated SB (green) sample. The width of each profile represents the error on the mean ($\sigma/\sqrt{N}$, where N is the number of spaxels in a radial bin), and the dashed line shows the bootstrap error estimation (see Appendix \ref{app_err}). \dsigsfr$=0$ is shown with a dashed line, and any values above this line indicate an enhancement in \sigsfr compared to the rSFMS. Post-merger galaxies consistently exhibit central \dsigsfr values that are twice as large as those of the isolated SBs.}
    \label{fig:profiles}
\end{figure*}

\section{Results}
\label{Sec: Results}

The main goal of this work is to discern whether or not post-merger galaxies have different \dsigsfr distributions from those of well-matched isolated galaxies with similarly high $\Delta$SFR. Given that the radial behavior of \dsigsfr will vary drastically depending on whether a galaxy is above, on, or below the SFMS, we divide our sample into bins of $\Delta\mathrm{SFR}$ \citep{Ellison2018StarMaNGA, Wang2019OnGalaxies, Sanchez2020SpatiallyGalaxies}. Galaxies are sorted into three bins of $\Delta\mathrm{SFR}$ to represent those on ($0<\Delta\mathrm{SFR}<0.25$), moderately above ($0.25<\Delta\mathrm{SFR}<0.5$), or drastically above ($\Delta\mathrm{SFR}>0.5$) the global SFMS. There are 70, 63, and 17 post-merger galaxies in these three categories, respectively, and 1177, 440, and 59 isolated SBs. To find differences in the average behavior between these post-merger galaxies and their matched isolated SBs, we find it best to compare the azimuthally averaged radial profiles constructed from all post-merger spaxels within a $\Delta\mathrm{SFR}$ bin. This helps to account for many patchy \dsigsfr profiles in some galaxies, which can lose spaxels particularly in the center and outskirts given our strict star-forming criteria. The median \dsigsfr profile for the post-merger galaxies and matched isolated SBs are shown in Figure \ref{fig:profiles}, with the post-merger galaxies shown in red and the isolated SBs in green. The error on the mean is shown as the width of each profile, and the bootstrap error in each radial bin is shown as a dashed line (see Appendix \ref{app_err} for bootstrap error estimation).

Figure \ref{fig:profiles} demonstrates that isolated SBs have relatively flat radial profiles in all three $\Delta\mathrm{SFR}$ bins, with only a modest negative gradient. As expected, based on the sample construction, the typical value of  the \dsigsfr in isolated SBs increases with global $\Delta\mathrm{SFR}$. For example, in the lowest $\Delta\mathrm{SFR}$ bin (representing galaxies within the upper end of the 1$\sigma$ MS scatter), the median \dsigsfr$\sim$ 0, whereas in the highest $\Delta\mathrm{SFR}$ bin the average spaxel is enhanced by almost a factor of two in \dsigsfr.  The mergers show a strikingly different behavior. In all three $\Delta\mathrm{SFR}$ bins, there is a much steeper \dsigsfr profile in the inner regions, indicative of a strong peak in SFR within $\sim$ 0.3 - 0.5 $R/R_e$. Therefore, whereas isolated SBs reach their elevated status by globally enhancing their star formation, mergers preferentially boost their star formation in their central regions.

The averaged galaxy profiles displayed in Figure \ref{fig:profiles} are enlightening, but we can also consider the individual galaxy profiles because mergers are known to induce significant galaxy-to-galaxy variations (e.g., \citealt{Thorp2019SpatiallyMaNGA,Thorp2022TheFormation}). Doing so also helps us account for the fact that some mergers will contribute more spaxels to the average radial profile than others. To quantify how often post-merger galaxies diverge from the general trends seen in Figure \ref{fig:profiles}, we compute the median \dsigsfr within 0.25$R_e$ for each post-merger galaxy and isolated SB, and then take the difference between the two. Doing so provides a numerical determination of how many isolated SB profiles are actually above the post-merger galaxy profile within 0.25$R_e$ (the opposite of what we would expect given the trends in Figure \ref{fig:profiles}). For the bins $0<\Delta\mathrm{SFR}<0.25$, $0.25<\Delta\mathrm{SFR}<0.5$, and $\Delta\mathrm{SFR}>0.5,$ we find 17\%, 33\%, and 0\% of the isolated SBs have larger \dsigsfr in their center than their assigned post-merger, respectively. Not only does this confirm that the difference in central starburst strength is real between post-merger galaxies and noninteracting galaxies, but it demonstrates how such a difference is more likely as we go to larger $\Delta\mathrm{SFR}$ values. For $0.25<\Delta\mathrm{SFR}<0.5,$ one-third of isolated SBs will have stronger central star formation activity than a matched post-merger galaxy. However, when we consider post-merger galaxies with $\Delta\mathrm{SFR}>0.5$, no isolated SBs exceed the merger-triggered central starburst. Thus, post-merger galaxies with heightened global SFRs are more likely to foster an extreme star-forming environment unique to the interaction process.

\section{Discussion and conclusion}
\label{Sec: Discussion}

The results of this work make clear that merger-induced starbursts are distinct from their isolated counterparts, exhibiting much larger central \dsigsfr independent of the global SFR enhancement. For mergers with moderate SFR enhancement ($0<\Delta\mathrm{SFR}<0.5$), some isolated SBs can exceed their matched post-merger galaxy in central \dsigsfr values. What leads to these exceptions remains uncertain. Stellar bars seem an obvious culprit; as mentioned in Sect. 1, they can create heightened nuclear star formation. To investigate a scenario where bars lead to post-merger-galaxy-like \dsigsfr profiles, we identified barred galaxies in our sample using the Galaxy Zoo classifications available for 96\% of MaNGA DR17 \citep{Willett2013GalaxySurvey,Hart2016GalaxyBias, Geron2023GalaxyGalaxies}. Using the criteria established in \cite{Willett2013GalaxySurvey}, we identified barred galaxies in our sample such that for ten or more participants, $>$43\% identified a feature or disk, $>$71.5\% voted the galaxy is not edge on, and $>$20\% identified a bar. Such conservative criteria results in 73 isolated SBs and 5 post-merger galaxies classified as barred by Galaxy Zoo volunteers (3.4\% and 4.5\% of each sample, respectively). However, only 30 of the isolated SBs with \dsigsfr profiles exceeding their matched post-merger galaxy are barred (4.2\%, a little less than the full isolated SB sample). Thus, we are fairly confident that the presence of a bar is not the primary cause of these exceptions. 

It is also possible that the isolated SB sample is contaminated with merging galaxies with features too faint to be identified by our visual classification of SDSS images. To approximate this contamination, we employ the recently published galaxy merger catalog from \citep{Ferreira2024GalaxyClassification}, which uses machine learning to identify galaxy mergers in the Canada-France Imaging Survey (CFIS) from the Ultraviolet Near Infrared Optical Northern Survey (UNIONS). UNIONS uses the Canada France Hawaii telescope to image galaxies at an $r$-band depth of 28.4 mag/arcsec$^2$, capturing much fainter features than the 22.7 mag/arcsec$^2$ r-band depth of SDSS. Of the 1676 isolated SBs, only 707 match the Right Ascension and Declination of an SDSS target in UNIONS within 2 arcseconds (an expected overlap given the minimum Declination in UNIONS is $\delta>$ 30\textdegree). Of those 707, only 21 are either pre- or post-coalescence objects in the merger catalog \citep{Ferreira2024GalaxyClassification}, meaning there is a possible 2.97\% merger contamination in the isolated SB sample. This small contamination fraction cannot explain the one-third of isolated SBs with greater central \dsigsfr than their matched post-merger galaxy.

It is more likely that the exceptional isolated SBs are caused by a combination of factors, including some properties not available for this sample, such as the gas properties of the MaNGA SBs. Observational connections between clumpy disks and increased star formation efficiency in isolated SBs indeed suggest that disk instabilities impact how gas is converted to stars \citep{Faisst2024COSMOS-Web:Z4}. It is also possible that post-merger galaxies have uniquely strong central star formation as a result of a heightened conversion of atomic gas to molecular gas \citep{Violino2018GalaxyMergers,Lisenfeld2019COFormation,Yu2024COSurvey}, leading to star formation efficiency increasing simultaneously with SFR \citep{Pan2018SDSSProperties}.
A galaxy-wide shock could induce this rapid transformation of the gas reservoir on a global scale, or cloud--cloud collisions could lead to greater pressure on giant molecular clouds, leading to localized enhancements \citep{Kaneko2017PropertiesFraction}. Indeed, simulations of interactions that create a starburst at coalescence predict that post-merger galaxies will have drastically different molecular cloud properties compared to regular star-forming galaxies, with clouds that are less gravitationally bound but simultaneously have shorter depletion times \citep{He2023MolecularMergers}. In such a scenario, there must be forces beyond self-gravity causing giant molecular clouds to collapse at a more rapid rate in a merger-induced starburst. Our work would imply that these extra forces are most pivotal in the center of the post-merger system, with the central starburst leading to a snowball effect where clouds collapse more rapidly, more stars form, and feedback from those stars further influences the central cloud population.

An investigation of giant molecular cloud properties in mergers would help confirm such a scenario, and would also help us understand why the central regions of mergers are so distinct. Case studies of nearby interacting galaxies like the Antennae system provide observational hints that mergers create unique conditions for giant molecular clouds \citep{Ueda2012UNVEILINGOBSERVATIONS, Wei2012TWOGALAXIES, Leroy2016APROCESSES, Brunetti2024Cloud-Scale3256}. However, a more comprehensive sampling of molecular cloud populations in mergers is necessary to create fair comparisons to recent cloud-scale surveys of ``normal'' star-forming galaxies, which have been examined in much more detail for a variety of galaxy properties \citep{Sun2018Cloud-scaleGalaxies,Sun2022MolecularPerspective}.

\begin{acknowledgements}

This project makes use of the MaNGA-Pipe3D dataproducts. We thank the IA-UNAM MaNGA team for creating this catalog, and the Conacyt Project CB-285080 for supporting them.

Funding for the Sloan Digital Sky Survey IV has been provided by the Alfred P. Sloan Foundation, the U.S. Department of Energy Office of Science, and the Participating Institutions. SDSS-IV acknowledges support and resources from the Center for High Performance Computing  at the University of Utah. The SDSS website is www.sdss4.org. SDSS-IV is managed by the Astrophysical Research Consortium for the Participating Institutions of the SDSS Collaboration including the Brazilian Participation Group, the Carnegie Institution for Science, Carnegie Mellon University, Center for Astrophysics | Harvard \& Smithsonian, the Chilean Participation Group, the French Participation Group, Instituto de Astrof\'isica de Canarias, The Johns Hopkins University, Kavli Institute for the Physics and Mathematics of the Universe (IPMU) / University of Tokyo, the Korean Participation Group, Lawrence Berkeley National Laboratory, Leibniz Institut f\"ur Astrophysik Potsdam (AIP),  Max-Planck-Institut f\"ur Astronomie (MPIA Heidelberg), Max-Planck-Institut f\"ur Astrophysik (MPA Garching), Max-Planck-Institut f\"ur Extraterrestrische Physik (MPE), National Astronomical Observatories of China, New Mexico State University, New York University, University of Notre Dame, Observat\'ario Nacional / MCTI, The Ohio State University, Pennsylvania State University, Shanghai Astronomical Observatory, United Kingdom Participation Group, Universidad Nacional Aut\'onoma de M\'exico, University of Arizona, University of Colorado Boulder, University of Oxford, University of Portsmouth, University of Utah, University of Virginia, University of Washington, University of Wisconsin, Vanderbilt University, and Yale University.

A.G. gratefully acknowledges support from the Valerie Kuehne Undergraduate Research Award (VKURA) from the University of Victoria.

M.T. thanks Leonardo Ferreira for advice on the UNIONS/MaNGA overlap and identification of merger contaminants in isolated SB sample, as well as the anonymous referee for important feedback on the initial draft.
\end{acknowledgements}



\bibliographystyle{aa} 
\bibliography{references} 
      

\begin{appendix} 
\label{app}
\section{Post-Merger and isolated galaxy distinction}
\label{app_merger}
\begin{figure*}
\centering
\includegraphics[width=0.75\textwidth]{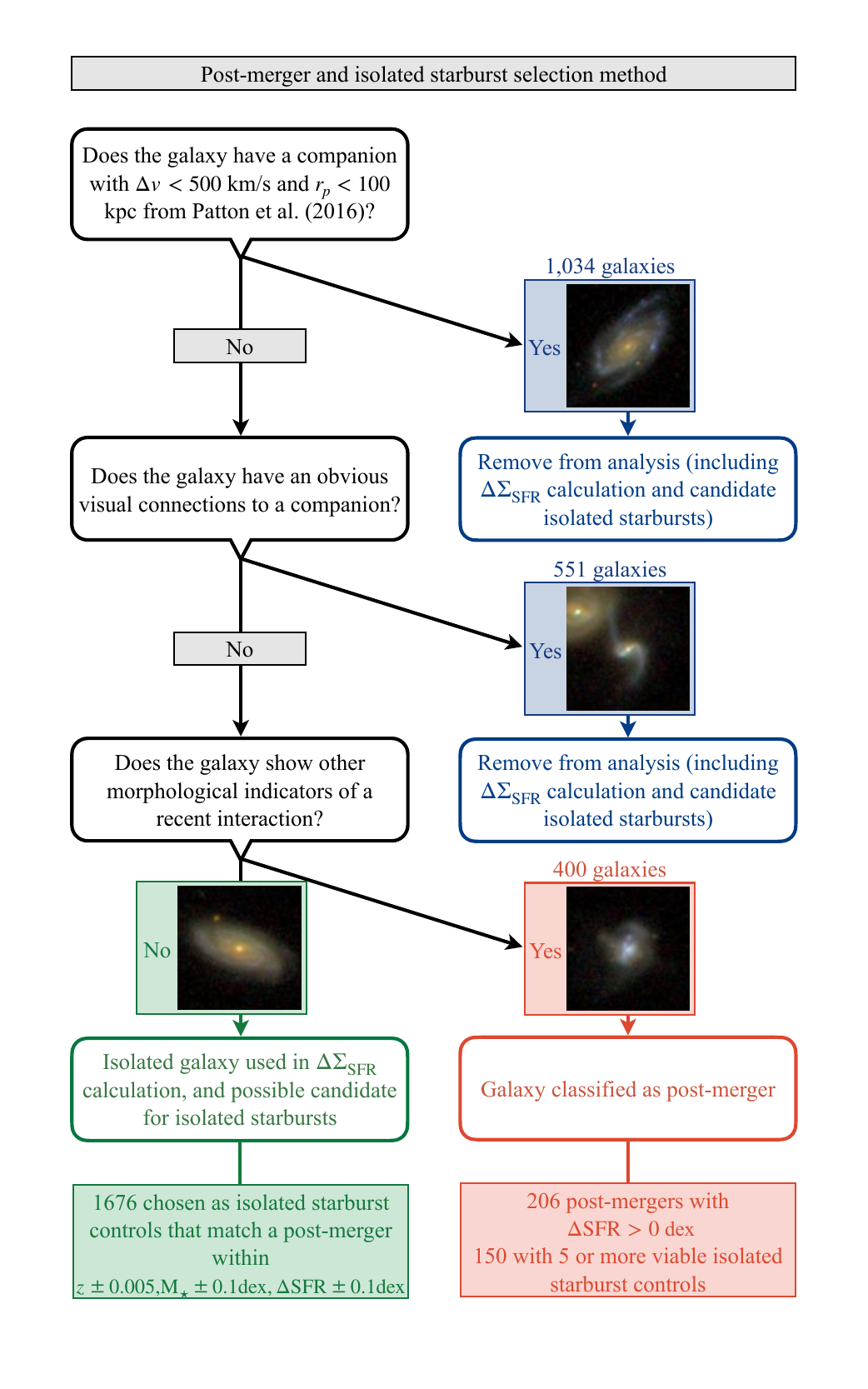}
\caption{Decision tree for both the isolated SB and post-merger sample, with SDSS {\it gri}-images provided as examples for each classification. 1034 interacting galaxies from the \cite{Patton2016GalaxySeparations} pair catalog (based on cuts in projected separation ($r_p$) and velocity difference ($\Delta v$)), and 551 visually identified interacting galaxies, are removed entirely from the analysis. The 400 post-merger galaxies which are visually identified are also removed from the pool of control spaxels used to compute \dsigsfrend. Only isolated galaxies are used in the \dsigsfr calculation. Of the 400 post-merger galaxies, 206 have $\Delta\mathrm{SFR}>$ 0 dex to be considered for this study of SBs, and a further 56 are removed for not having five or more viable isolated SB controls.}
\label{fig:decision_tree}
\end{figure*}

To provide further clarification and easy understanding of the merger selection method described in Section \ref{Sec: Merger Sample}, we additionally include decision tree that summarizes the post-merger and isolated SB selection process in Figure \ref{fig:decision_tree}. Additional information about the number of galaxies classified as either spectroscopically or visually confirmed interacting pairs is provided, as these are removed entirely from the analysis presented in this work. After isolated and post-merger galaxies have been differentiated, both based on visual classification and overlap with the \cite{Patton2016GalaxySeparations} galaxy pair catalog, the post-merger SB sample is selected such that all post-merger galaxies have $\Delta \mathrm{SFR}>$ 0 dex. Finally, isolated SBs are chosen as a comparison sample given they match a target post-merger in $z$, \Mtotend, and $\Delta \mathrm{SFR}$. This results in a sample of 150 post-merger galaxies that have at least five viable controls, and the 1676 isolated SBs that constitute that control sample.

\section{Properties of post-merger sample and the number of comparable isolated SBs}
\label{app_control}

\begin{figure*}
    \includegraphics[width=0.8\textwidth]{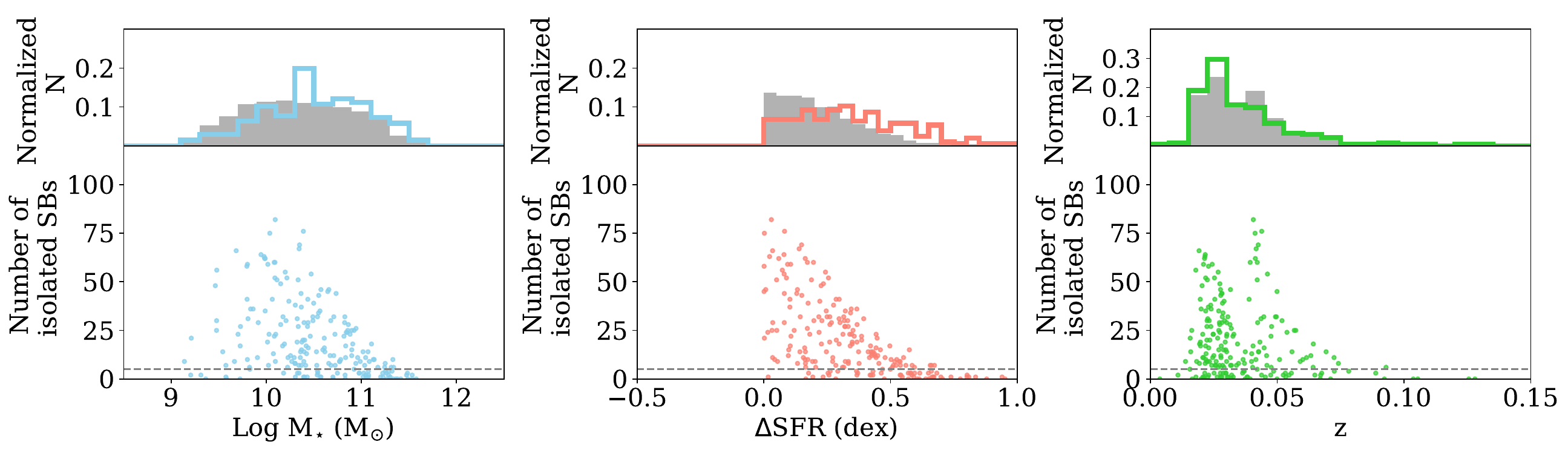}
    \centering
    \caption{\textit{Top:} Histograms of the normalized galaxy count for the isolated (solid) and post-merger (outline) SBs in terms of \Mtot (left), $\Delta\mathrm{SFR}$ (middle), and $z$ (right). Overall, the post-merger galaxies on average have slightly higher \Mtot and $\Delta\mathrm{SFR}$, appropriate for a scenario where mergers gain mass in an interaction and trigger a starburst. \textit{Bottom:} The number of isolated SBs matched to each post-merger as a function of the post-merger's \Mtot (left), $\Delta\mathrm{SFR}$ (middle), and $z$ (right). The minimum number of isolated SB controls required for our analysis (5) is marked with a gray dashed line. For all three matched variables, the number of matched isolated SBs decreases as the matched variable increases.}
    \label{fig:Control}
\end{figure*}

To understand why a quarter of the post-merger sample fail to have five or more well-matched isolated SBs, we investigate the matching properties of the post-merger and isolated SB sample in Figure \ref{fig:Control}. The top row displays the distributions of \Mtotend, $\Delta\mathrm{SFR}$, and $z$ for all galaxies with $\Delta\mathrm{SFR}>0$ dex, separating the 206 post-merger galaxies (colored) and 2113 isolated SBs considered as candidate control galaxies (gray). post-merger galaxies on average have larger \Mtot and $\Delta\mathrm{SFR}$ than isolated SBs, leading to a lack of viable comparison SBs for the most massive, most star-forming post-merger galaxies. Indeed, the number of isolated SBs that match each post-merger within our criteria decreases with \Mtot and $\Delta\mathrm{SFR}$, as is demonstrated in the bottom row of Figure \ref{fig:Control}. Thus the most extreme post-merger SBs ($\Delta\mathrm{SFR}>0.7$ dex) are excluded from our analysis as they fail to acquire 5 isolated SB control galaxies.

\section{Uncertainty in radial profiles}
\label{app_err}

To estimate the uncertainty in the median \dsigsfr profiles, we employ a bootstrap error estimation with the following method: within each $\Delta\mathrm{SFR}$ bin a random number of post-merger galaxies are selected with replacement such that the previous sample size is achieved (70 in 0 dex $<\Delta\mathrm{SFR}<$ 0.25 dex, 63 in 0.25 dex $<\Delta\mathrm{SFR}<$ 0.5 dex, 17 in $\Delta\mathrm{SFR}>$ 0.5 dex), but there may be repeats of post-merger galaxies while others are left out entirely. The subsequent isolated SBs for this `new' post-merger set are selected as well, using the same selection criteria as the original analysis. The median \dsigsfr profile for the new post-merger and isolated SB samples are then created, with slight variations to that shown in Figure \ref{fig:profiles}. Randomized profiles are created 1000 times, and over each iteration we calculate the difference from the original median \dsigsfr value in each radial bin. We then compute the median difference over those 1000 iterations to quantify the uncertainty in the profile. The median difference quantifies a bootstrap error for each radial bin, and is shown as a dashed line in Figure \ref{fig:profiles}. The difference in post-merger and isolated SB profiles being greater than the bootstrap error assures that these results are not sensitive to the underlying galaxy sample and the galaxy-to-galaxy variations expected for \dsigsfr profiles in MaNGA.

As can be seen in Figure \ref{fig:profiles}, the bootstrap error is similar to the error on the mean for each profile, though at times slightly larger. Both are significantly larger than the error in \sigsfrend, which can be computed for each spaxel from the error on the H$\alpha$ emission provided by \pipedend. Given we employ a conservative S/N$>$3 criterion for H$\alpha$, the average \sigsfr error based on the emission uncertainty is 0.008 dex for the SB sample (far less than the dex changes measured for \dsigsfr profiles in this work). Thus the uncertainty in Figure \ref{fig:profiles} is better quantified by the galaxy-to-galaxy variation, which creates far more scatter in the median radial profiles.

\begin{figure*}
\centering
\includegraphics[width=0.75\textwidth]{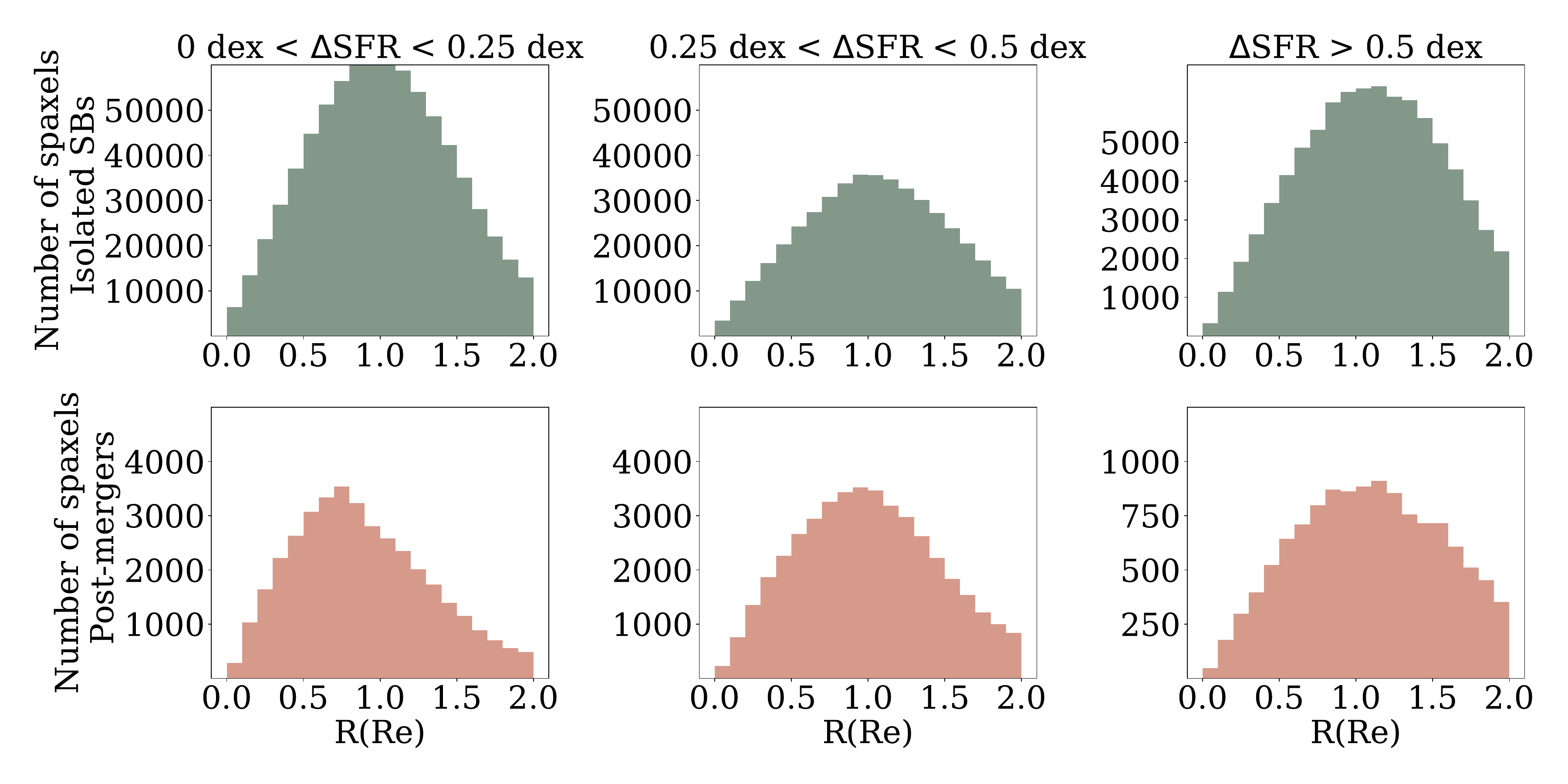}
\caption{The number of spaxels in each radial bin shown in Figure \ref{fig:profiles}, with the isolated SBs (top) and post-merger galaxies (bottom) shown separately to account for the significant increase in spaxel count in the isolated SB comparison sample. The least populated radial bin is always $R<0.1R_e$ as expected; larger radial bins will contain more spaxels than inner bins. Beyond $R=1R_e$, the quality cuts we employ to select high S/N star-forming spaxels result in less spaxels contributing (given H$\alpha$ S/N falls off with radius).}
\label{fig:spax_count}
\end{figure*}

\begin{figure*}
\centering
\includegraphics[width=0.75\textwidth]{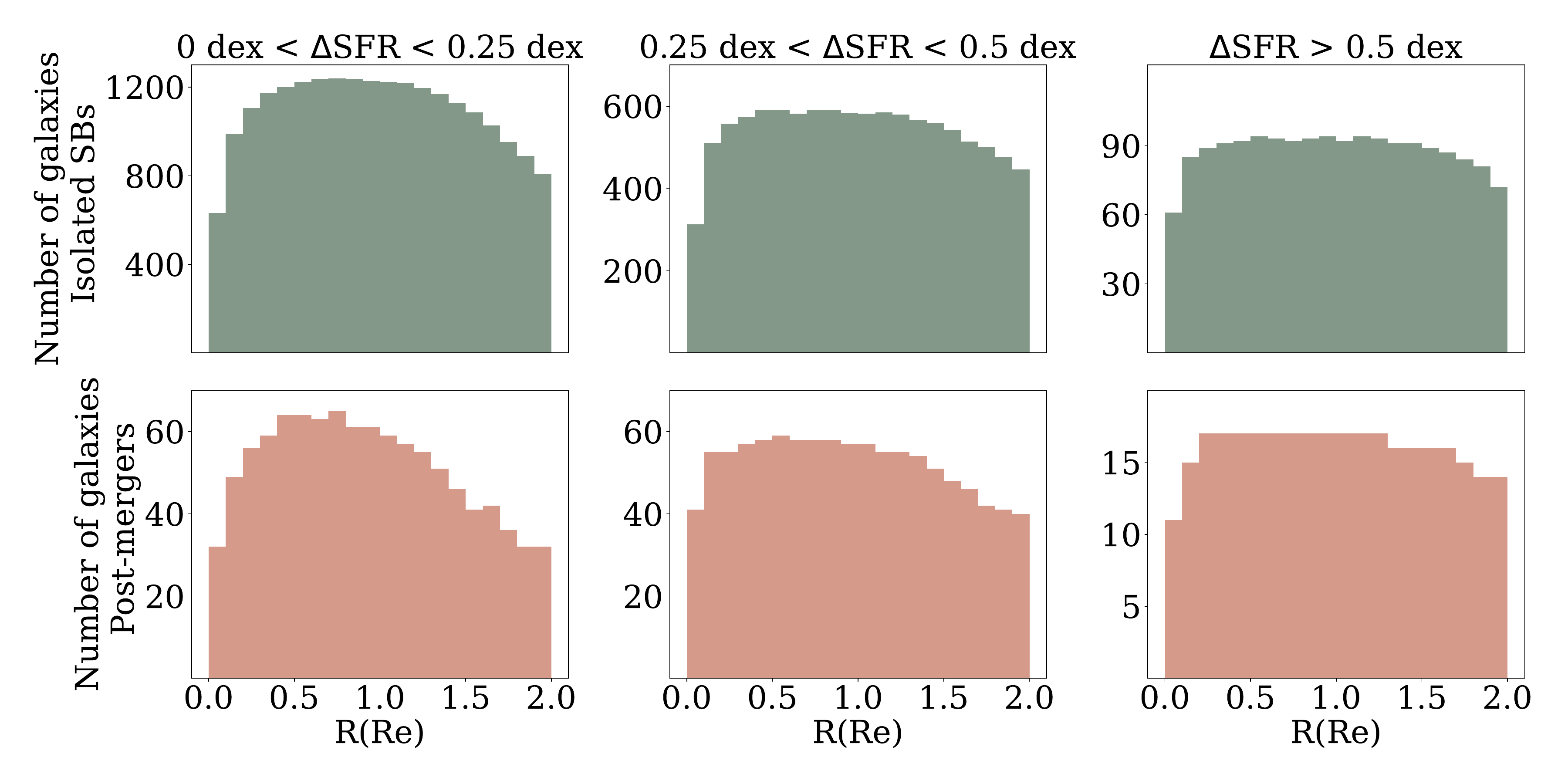}
\caption{The number of galaxies in each radial bin shown in Figure \ref{fig:profiles}, with the isolated SBs (top) and post-merger galaxies (bottom) shown separately to account for the significant increase in galaxy count in the isolated SB comparison sample. Note that even in the least populated bin ($R<0R_e$) more than 50\% of the galaxies in the respective $\Delta\mathrm{SFR}$ bin contribute.}
\label{fig:gal_count}
\end{figure*}

We also consider how the number of galaxies and spaxels that contribute to each radial bin may impact the differences in isolated SBs and post-merger galaxies at small radii.
Figure \ref{fig:spax_count} and Figure \ref{fig:gal_count} show  the number of spaxels and galaxies that contribute to each radial bin for the profiles shown in Figure \ref{fig:profiles}, respectively. The smallest and largest radial bins contain the least number of spaxels. The former is expected given more spaxels (regardless of meeting our quality cuts) are present at larger radii than smaller radii. The lack of spaxels at large radii is  mainly a result of spaxels failing to meet the S/N$>$3 restriction on \dsigsfr measurements. The galaxy contribution per radial bin is more uniform, with the least number of galaxies contributing to the smallest radial bin ($R<0.1R_e$). However, within $R<0.1R_e$ 50\% or more of galaxies within the respective $\Delta\mathrm{SFR}$ bin contribute to the \dsigsfr profile (i.e., 17 post-merger galaxies contribute to $\Delta \mathrm{SFR}>0.5$ dex, and 10 of those contribute to $R<0.1R_e$). Therefore, we are confident our results concerning the central \dsigsfr behavior are not being driven by a handful of galaxies or bin sample sizes.

\end{appendix}

\end{document}